\documentclass[letterpaper, 10 pt, conference]{ieeeconf}  

\IEEEoverridecommandlockouts                              
\usepackage{booktabs}

\title{\LARGE \bf
Pareto Optimal Re-ranking with Semi-Automated Content Credibility Detection
}

\usepackage{amsmath,amsfonts,amssymb}

\newcommand{\consumed}{K}

\newcommand{\distsf}{D_{\textrm{SF}}}

\DeclareMathOperator{\idperm}{\mathbf{id}}
\DeclareMathOperator{\Tr}{Tr}

\usepackage{hyperref}
\usepackage{listings}
\usepackage{graphicx}

\usepackage{xcolor}

\author{Yigit Ege Bayiz, Arash Amini, and Ufuk Topcu
\thanks{Yigit Ege Bayiz is with Faculty of Electrical and Computer Engineering
        University of Texas at Austin,
        {\tt\small egebayiz@utexas.edu}}%
\thanks{Arash Amini, and Ufuk Topcu are with the Oden Institute for Computational Engineering and Sciences, University of Texas at Austin, 
        {\tt\small a.amini@utexas.edu}, {\tt\small utopcu@utexas.edu}}%
}

\begin{document}

\maketitle
\thispagestyle{empty}
\pagestyle{empty}

\begin{abstract}
Social media posts often include misinformative or misleading content, diminishing the expected credibility of content feeds. We present an optimization-based method to improve the credibility of news content on social media feeds by refining existing content rankings. This method is based on a dual-objective optimization approach that minimizes the Spearman's footrule distance to the original ranking to maintain the original content order while incorporating an additional linear cost objective to elevate the expected credibility of the content feed. Additionally, we propose a robust semi-automated pipeline for assigning credibility scores to content based on a mixture of retrieval-augmented score assignments and human-generated fact-checks. This semi-automated pipeline helps ground the credibility assignment using human-generated labels while ensuring the algorithm extends to posts with few or no human-generated labels. We showcase our approach through an experimental setup using real-world data collected over X (Twitter), where we assign the credibility scores based on a mixture of user-generated community notes and retrieval augmented generation. The method we present leads to at most $7\%$ deviation in both optimization objectives from the Pareto optimal front with known initial ranking values. Additionally, the algorithm allows for incorporating different measures for source credibility, making it applicable across various social media platforms.

\end{abstract}

\section{Introduction}\label{sec1}

The propagation of misinformative news threatens to undermine political and social discourse. 
Uncredible and misleading news manipulates public opinion through misinformation, shaping political and social outcomes. The lack of regulations on information circulating across social media creates an open arena where uncredible news can spread freely.
Recent studies indicate that over $50\%$ of U.S. adults occasionally rely on social media sites for their daily news, with over $17\%$ of U.S. adults \textit{regularly} using social media for news consumption \cite{PewResearch-2023-FactSheet}.


Many social media websites, such as X (Twitter), Reddit, LinkedIn, or Facebook, organize their content in a continuous \textit{feed} the reader can scroll through to see relevant content sequentially. The social media platforms use \textit{ranking algorithms} to determine the order of the content in the user's feed. Most existing ranking algorithms in social media prioritize engagement metrics, such as user ratings, click probability, and engagement time \cite{vinod2023survey}. Optimizing the ranking based on engagement metrics is profitable for social media websites, as it simultaneously improves the user experience by omitting content that is uninteresting to the user---thus retaining more users---and maximizes the time users spend on the social media website, providing more advertisement opportunities \cite{huszar2022amplification, bak-coleman2022combining}. 

However, only optimizing for user engagement can be detrimental to public benefit \cite{wu2017optimizinglongterm}, as it creates a fertile ground for sensationalized and uncredible news to flourish. The credibility of news has little effect on users' probability of sharing \cite{Pennycook2021shifting}, which allows uncredible news to appear frequently in users' content feeds, propagating rapidly between the users. Ranking algorithms that take the credibility of the content source into account exist. However, such algorithms are not cost-effective for social media platforms to implement as they require fine-tuning the existing ranking models or training a new ranking model from scratch. Additionally, fine-tuning the existing models is only feasible from the platform's end-point, which can introduce new biases into the feed.

We address these challenges by developing a re-ranking method that improves an existing content feed by re-ordering elements to enhance the credibility of the content seen by the user while maintaining user engagement levels. This method poses re-ranking as a dual-objective optimization problem of simultaneously minimizing Spearman's footrule \cite{diaconis1977spearman} distance to preserve the original ranking while maximizing a linear cost function reflecting the source credibilities. This formulation allows for the balanced incorporation of credibility without disrupting the user experience. We solve this problem using a shortest augmenting path algorithm \cite{Crouse2016assignment} by showing its equivalence to a bipartite matching problem. 

A major advantage of the proposed method is that it does not require knowledge of the ranking algorithm used to generate the original content order. Instead, it seamlessly adapts to existing ranking algorithms by treating them as black-box systems and directly refining their outputs. This independence from the original ranking scores makes the method a candidate for client-side re-ranking of the content---for instance, a browser extension---that can adapt the ordering of the content feed for the end user after receiving the ranked content list from the social media website. Said decentralized re-ranking is also permissive for the credibility scores to be tailored to the end user without jeopardizing privacy, by relying on locally stored data.


We showcase the real-world applicability of the proposed method through a setup utilizing X's data from 2021 to 2024. We assign credibility scores to content by combining X’s \textit{community notes}\footnote{Accessible from https://communitynotes.x.com} data with a retrieval-augmented generation model. Specifically, we measure the credibility of the content feed using a novel semi-automated pipeline that integrates human-generated labels from community cotes with automated labels derived from a retrieval-augmented generation model based on the knowledge base from AskNews\footnote{Accessible from https://asknews.app/en/about}. This integration ensures reliable credibility scoring, even in scenarios where human-generated labels are insufficient. At the same time, it prevents automated labels from dominating the scoring process when enough human-generated labels are available, maintaining a democratic credibility evaluation.

The primary contributions of this paper are as follows.
\begin{enumerate}
    \item We formulate and solve a re-ranking problem for reducing users' misinformation exposure by optimizing for the expected credibility of the content feed.
    \item We provide a pipeline to apply the proposed re-ranking on X using a novel mixture of community notes and retrieval augmented generation to assign credibility scores.
    \item We analyze the Pareto optimality tradeoff associated with the proposed pipeline over real data from X.
\end{enumerate}

\section{Related Work}\label{sec2}
\subsection{Misinformation Exposure}

Understanding users' engagement with misinformative content requires insights into their news consumption habits \cite{ahmed2022social}. Although natural language processing techniques, combined with network theory, have achieved remarkable accuracy in detecting online misinformation \cite{comito2023multimodal,aimeur2023fake}, these models are yet to be deployed for real-time detection of online misinformation on a large scale. However, misinformation narratives often subtly aim to influence users' beliefs over time \cite{o2019misinformation,molina2021fake}. These techniques also depend on partially misleading information, which may not be false in isolation but becomes misinformation when viewed within context \cite{pillai2021effects}. For example, many misinformation campaigns seek to erode trust between the public and government institutions to increase the public's susceptibility \cite{stubenvoll2021media,nahum2021erosion}.


Many social media platforms, like X and Facebook, attempt to reduce exposure by filtering specific types of content from users' feeds. While effective, this method often overlooks posts containing partial misinformation or content not flagged for filtering. Additionally, such filtering can raise concerns about free speech in certain regions. Alternatively, re-ranking content based on the credibility of the post or the user sharing it can effectively lower users' exposure to misinformation while avoiding censorship. Introducing credibility into ranking systems would also incentivize media outlets, influencers, and political elites to improve reliability by sharing more trustworthy information to maintain visibility. 

\subsection{Rank Aggregation}
Rank aggregation is a well-established research area that focuses on combining multiple ranked lists into a single consolidated ranking. Broadly, rank aggregation methods fall into two categories: score-based aggregation methods assign numerical scores to items, reflecting their importance or relevance across different rankings. Condorcet's method \cite{de2014essai, Young_1988} and additive score aggregation fall under this category. All of these methods produce a final ranking by averaging the scores of each ranking algorithm to construct new scores that provide a trade-off between individual rankings. Score-based approaches are effective when individual rankings provide explicit scores.

Rank-based aggregation methods focus on the relative positions of items within each ranking rather than numerical scores. Dwork et al. \cite{dwork2001aggregation} analyzes several of these rank aggregation methods including the traditional Kemeny-Young \cite{kemeny1959mathematics} vote aggregation method, which minimizes Kendall's tau \cite{kendall1939tau} distance to each ranking, and polynomial-time methods using Spearman's footrule \cite{diaconis1977spearman} distance to minimize disagreements between the aggregated ranking and the individual rankings. Fujita et al. \cite{Fujita2020unsupervised} introduce HPA, which first averages the rankings to create a pseudo-ranking, then evaluates each original ranking to filter out the low-performing ones before re-averaging. Other rank-based aggregation methods assign statistical scores to items based on their ranking positions and then rank the resulting scores to obtain an aggregate ranking.\cite{fox1993aggregation, Fagin2003EfficientSS, cormack2009reciprocal}. 

The re-ranking approach we propose in this paper enhances an existing content feed by reordering elements to improve source credibility while maintaining user engagement levels. Theoretically, the re-ranking step we utilize draws existing ranking aggregation methods. However, the proposed approach differs from traditional ranking aggregation methods by focusing on refining a single ranking with additional score optimization criteria rather than combining multiple rankings. Therefore, it is more accurate to refer to it as a re-ranking method, rather than an aggregation method. Additionally, we provide a detailed description of the measurement of credibility scores.

\section{Problem Statement}\label{sec3}
We define a \textit{content feed}, as a system in which the user sequentially interacts with an ordered list of contents, which we call \textit{posts}. The content feed contains a batch of $\consumed$ posts, ordered by a ranking algorithm. We call this ordering the \textit{canonical ordering}. If the user reaches the end of the content feed, the ranking algorithm prepares a new batch of posts.

Let $q_p$ denote the probability of the user seeing the $p$'th post in a batch. We assume $q_p$ is non-increasing in $p$ since the user interacts with the content sequentially. We model $q_p$ as exponentially decaying with base $\lambda$. That is,
\begin{equation}
    q_p = \lambda^{p}.
\end{equation}
The above definition for $q_p$ is equivalent to assuming that the user has a fixed probability $(1-\lambda)$ of stopping interacting with the feed at each post $p$. The problem formulation generalizes to arbitrary user models so long as $q_p$ is a non-increasing function of $p$.


We define the credibility $c_p$ of a post $p$ as the probability that a random human rater deems the post as being non-misinformative. The definition we use in this paper serves as a stipulation to establish a workable ground truth for analysis. For a more rigorous definition of credibility, let $ Z_p \sim \text{Bernoulli}(c_p) $ be a Bernoulli random variable representing the label assigned by a random human fact-checker to a given post $p$, where $ Z_p = 1 $ if the post is not misinformative and $Z_p = 0$ if it is misinformative. We then define the post $p$'s credibility as the expectation $\mathbb{E}[Z_p] = c_p$.

The goal of re-ranking is to refine the canonical ordering to improve the expectation of the total credibility score of the posts seen by the user. Suppose that canonical ordering optimizes some value function $V$, called the \textit{original value}, which assigns a score to all permutations $\sigma\in S_\consumed$ of the content feed. In this case, re-ranking simplifies into the following dual-objective optimization problem with a weight parameter $\alpha$,

\begin{align}\label{mainprob1}
    \max_{\sigma \in S_\consumed} \quad& \alpha V(\sigma) + \sum_{p=1}^{\consumed} q_p c_{\sigma(p)}.
\end{align}

The above problem is an example of score-based ranking aggregation, and it is well-studied in the literature \cite{ijcai2024p915}. In the case the original value function itself is an item-wise expected cost, that is
\begin{align}
    V(\sigma) = \sum_{p=1}^{\consumed} q_p v
    ({\sigma(p)}),
\end{align}
where $v_{p}$ is the value of post $p$, then any Pareto optimal solution of the problem is trivially solvable by finding the permutation $\sigma$ that sorts $v_{\sigma(p)} + \alpha c_{\sigma(p)}$ in descending order. Here, $\alpha \geq 0$ is a tunable parameter that parametrizes the set of all Pareto optimal solutions.

In this paper, we consider the scenarios, where the original values $V$ are inaccessible. In some cases, the original value function, may not exist explicitly, such is the case if the canonical ordering is an aggregation of multiple orderings. And in cases where the value function may exist, the re-ranking algorithm may not have access to it. For instance, if the re-ranking algorithm runs on the client-side or a third-party application, it is in general not possible for it to access the original value $V$. 

We focus on ranking-based aggregation as a surrogate problem to \ref{mainprob1}. Let $S_\consumed$ denote the space of permutations acting on the batch of $\consumed$ contents. We associate an intrinsic permutation cost based on Spearman's footrule distance $\distsf(\sigma, \idperm)$ with each permutation $\sigma \in S_\consumed$. 

\begin{align}\label{mainprob}
    \min_{\sigma \in S_\consumed} \quad& \alpha \distsf(\sigma, \idperm) - \sum_{p=1}^{\consumed} q_p c_{\sigma(p)},
\end{align}
where $\alpha$ is a weight parameter that determines the relative importance of the optimization objectives. This problem maximizes the credibility of the content user views while simultaneously minimizing the distance between the canonical ordering and the permuted ordering. Thus reducing the misinformative content users are exposed to, while not violating the platform objectives significantly.

\section{Data}
\subsection{X Community Notes Data}
\begin{table}
    \centering
    \caption{Yearly distribution of the community notes data.}
    \begin{tabular}{l|ccc}
\toprule
\textbf{Year} & \textbf{Posts} & \textbf{Community Notes} & \textbf{Notes per Post} \\ 
\midrule
2021 & 14897 & 19231 & 1.29 \\ 
2022 & 15321 & 21636 & 1.41 \\ 
2023 & 241666 & 399237 & 1.65 \\ 
2024 & 442138 & 738792 & 1.67 \\ 
\midrule
Total & 714022 & 1178896 & 1.65 \\ 
\end{tabular}
    \label{tab:dataset_breakdown}
\end{table}


The experiments in this paper use Community Notes data from X to demonstrate the application of the proposed method on a real-world dataset. The Community Notes is a crowd-sourced \cite{wojcik2022birdwatch} program X employs to flag posts that contain misinformation or misleading content. It works based on allowing certain contributors to leave notes on posts, and ask users to rank how useful notes are. Table \ref{tab:dataset_breakdown} presents a yearly breakdown of the dataset. 

The data in Table \ref{tab:dataset_breakdown} represents the entire corpus of community notes data distributed by X. For the empirical studies, it was not feasible to pass all of the posts through the retrieval model to generate artificial labels for each note. Instead, we prepare a representative test set from the corpus by sampling $2400$ posts each year and their associated community notes. 


Among the $2400$ posts we sample each year, we sample half of them as the top $1200$ posts in terms of the number of community notes, and we sample the remaining half uniformly at random from the remaining posts. This split allows us to also analyze the improvements caused by using retrieval-augmented artificial rankings, as the impact of artificial rankings is significantly less in posts with many existing community notes.



\subsection{PolitiFact Data}\label{sec:politifact}

We evaluate the accuracy of the artificial labels on a dataset of claims and fact-checks from PolitiFact\footnote{Accessible from https://www.politifact.com}, a non-profit organization of expert journalists who provide accuracy assessments to claims and statements made by elected officials and others involved in U.S. politics. PolitiFact fact-checks provide a reliable grounding to assess the performance of automated fact-checking algorithms, and existing literature routinely uses them as a benchmark \cite{guo-etal-2022-survey, xiao2024challengesmachinelearningtrust}.
PolitiFact assigns one of six different labels to each claim depending on how misinformative the claim is. For the context of this paper, we enumerate these labels with numbers from $0$ to $5$ with increasing levels of credibility, That is, $0$ refers to misinformative claims and $5$ refers to completely true claims.

For our experiments, we collect $1165$ claims and corresponding fact checks from between January $2021$ and October $2024$, which coincide with the dates in the X community notes dataset. We construct the dataset by sampling $200$ claims corresponding to each label, with the exception of the label $5$ for which there were only $165$ in the time frame. This sampling allows us to construct a balanced dataset for the artificial misinformation labeling benchmarks.

\section{Methods}\label{sec4}

Based on the earlier discussion, maximizing credibility in social media feeds naturally splits into two subproblems: \textit{credibility assignment} and \textit{ranking refinement}. In this section, we describe the methods we use to tackle both of these problems. We tailor our methodology and nomenclature to the social media feed structure on X, as our subsequent experiments use data from X. However, the methods we present extend naturally to other social media platforms with similar content feeds despite possibly needing a slightly different implementation.

The summary of the entire processing pipeline is as follows. First, we generate the credibility scores for each post by aggregating scores from both human fact-checks, coming from user-generated \textit{community notes}, and automated fact-checks we acquire by querying a retrieval augmented generation model. Second, we alter the canonical ordering based on credibility scores using a re-ranking method. We detail each of these parts in the subsequent sections.


\subsection{Credibility Assignment}
In this section, we provide a method for estimating the credibility score $c_p$ for a given post $p$ using a mixture of both human-generated and automated fact-checks.
\subsubsection{Human-Generated Credibility Scores}
The implementation we present in this section is specific to the social media platform X, and we use the community notes feature of X heavily in our credibility assignment. However, similar structures, such as user comments and ratings either already exist or are implementable in other social media platforms. 

In X, community Notes are user-generated annotations on posts, containing binary labels that indicate whether a post is misinformative or not. Given a post $p$, we treat each community note $z_{p,i}$ as a sample of the Bernoulli random variable $Z_p$. Thus, using only community notes data, a simple frequentist estimator for the credibility $c_p$ is the sample mean of community notes $[z_{p,1},\dots,z_{p,n}]$. That is,
\begin{equation}
    \hat{c}_p^{\textrm{human}} = \frac{1}{n} \sum_{i=1}^{n} z_{p,i},
\end{equation}
where $n$ is the number of community notes for the post.

However, community notes require effort from users to create and are thus relatively sparse. More than $70\%$ half of the posts in the X dataset we use have less than $7$ community notes. To address this issue, utilize additional data from ratings assigned to each community note. Multiple ratings accompany each community note, labeled as either \texttt{HELPFUL} or \texttt{NOT\_HELPFUL}.

To reduce variance in the estimation of credibility, treat each \texttt{HELPFUL} rating as an additional community note agreeing with the original note's label $z_{p,i}$, and each \texttt{NOT\_HELPFUL} rating as an additional note disagreeing with $z_{p,i}$. This approach effectively increases the sample size by incorporating the ratings as weighted votes. Thus, the estimator becomes,
\begin{equation} \label{humancred}
    \hat{c}_p^{\textrm{human}} = \frac{\sum_{i=1}^{n} \left( z_{p,i} \cdot h_{p,i}^+ + (1 - z_{p,i}) \cdot h_{p,i}^- \right)}{\sum_{i=1}^{n} \left( h_{p,i}^+ + h_{p,i}^- \right)},
\end{equation}
where $h_{p,i}^+$ and $h_{p,i}^-$ are respectively the number of \texttt{HELPFUL} and  \texttt{NOT\_HELPFUL} ratings for community note $i$ of post $p$.

In the X dataset we use in the paper, incorporating the ratings reduces the standard deviation of the final credibility by approximately $10$ fold, significantly improving the reliability of measurements.

\subsubsection{Augmenting Credibility Estimation with Artificial Labels}\label{sec:artificiallabels}
The number of community notes per post in the dataset follows a geometric distribution, with the majority of posts having fewer than ten community notes. This sparsity makes relying solely on human-generated labels infeasible for accurate credibility estimation. To address this limitation, we augment the human-generated labels by incorporating artificially generated labels. We generate these artificial labels using a retrieval-augmented generation (RAG) pipeline that analyzes existing news sources and compares the post contents with these sources to generate a binary labeling, akin to a community note.

We use the news knowledge base from AskNews, a RAG-based news assistant tool, as the foundation for retrieval. The AskNews knowledge base contains over 50,000 news sources in more than 13 languages, providing a reliable knowledge base for misinformation detection. For each post $p$, we use the AskNews API to retrieve relevant news articles that correspond to the content of the post and input the retrieved articles, along with the original post and a predefined prompt, which we include in Appendix \ref{apx:prompting}, into a text-generation large language model (LLM) to generate an artificial label $z_{p,0}$ similar to a community note labeling. In most of our experiments, we use GPT4o by OpenAI \cite{OpenAI} due to its accuracy. We also test and compare to several alternative models in Section \ref{sec_ragtests}.

After generating the artificial label $z_{p,0}$ for each post, we assign fictitious positive $k^+$ and negative $k^-$ ratings to provide a baseline weight to these artificial community notes. These ratings behave as a measure of how reliable the artificial label is, similar to ratings on community notes. Note that these fictitious ratings are independent of the post, meaning that the weights of the artificial labels are identical across all posts.

We merge the artificial labels with the human-generated community notes to generate aggregated credibility scores as follows,
\begin{equation}
\footnotesize
\begin{split} \label{credscores}
    &\hat{c}_p =\\
    &\frac{ z_{p,0} \cdot k^+ + (1 - z_{p,0}) \cdot k^- +  \sum_{i=1}^{n} \left( z_{p,i} \cdot h_{p,i}^+ + (1 - z_{p,i}) \cdot h_{p,i}^- \right)}{\sum_{i=1}^{n} \left( h_{p,i}^+ + h_{p,i}^- \right) + k^+ + k^-}.
\end{split}
\end{equation}

\subsection{Re-ranking}
This section addresses the second subproblem, \textit{re-ranking}, in which the objective is to reorder the posts in a user's feed to maximize overall credibility while maintaining similarity to the canonical ranking provided by the platform. This approach aligns with the rank aggregation methods explored by Dwork et al. \cite{dwork2001aggregation} based on Spearman's footrule distance, which leads to a linear assignment problem. However, the formulation we use differs slightly from this existing approach. Thus, we describe it here briefly.

The general form problem (\ref{mainprob}) is a mixed integer optimization problem and might not have a general polynomial-time solution. However, in the special case of \label{mainprob2}, where the permutation norm is calculated using the Spearman's footrule \cite{diaconis1977spearman} metric, the problem reduces to a linear assignment problem.

The Spearman's footrule distance is defined by \cite{diaconis1977spearman}
\begin{equation}
	\distsf(\sigma) = \sum_{i=1}^{n} |i - \sigma(i)|.
\end{equation}
Let $\mathbf{\Pi}$ be the permutation matrix corresponding to $\sigma$, where $\mathbf{\Pi}_{i,j} = 1$ if $\sigma(j) = i$ and $0$ otherwise. Also let $\mathbf{L}_n$ is the matrix of absolute differences,
\begin{equation}
	\mathbf{L}_n = [\, |i - j| \, ]_{i,j}^{n},
\end{equation}
Then, have we have 
\begin{equation}
    \distsf(\sigma) = \Tr(\mathbf{\Pi} \mathbf{L}_n).
\end{equation}
This equality follows directly from by writing the sum formula for both the matrix multiplication and matrix trace and verifying it is equal to the definition of Spearman's footrule distance.

Let $\mathbf{q} = [q_p]_p$ and $\mathbf{c} = [\hat{c}_p]_p$. Then we have,
\begin{equation}\label{eq: costTrace}
    \sum_{i=1}^{\consumed} q_p c_{\sigma(p)} = (\mathbf{\Pi} \mathbf{c})^\top \mathbf{q} = \Tr(\mathbf{\Pi} (\mathbf{q}^\top \mathbf{c})).
\end{equation}
Thus, we can rewrite the optimization problem in equation \ref{mainprob} as,
\begin{align}\label{bimatchprob}
    \min_{\mathbf{\Pi}} \quad& \Tr(\mathbf{\Pi}(\mathbf{\alpha \mathbf{L}_n - \mathbf{q}^\top}\mathbf{c})),\\
    \textrm{s.t.} \quad& \mathbf{\Pi} \textrm{ is a permutation matrix.}
\end{align}

Intuitively, the above formulation defines a matrix of credibility costs $[-q_i c_j]_{i,j}$ associated with having content $j$ at position $i$. We then alter this matrix by the matrix $\alpha \mathbf{L}_n$ describing the cost of moving content $j$ to position $i$. Left multiplying with $\mathbf{\Pi}$ and then taking the trace equates to summing the elements of the cost matrix 
$\alpha \mathbf{L}_n - \mathbf{q}^\top $, which yields the total cost of permutation $\sigma$.

The optimization problem in equation \ref{bimatchprob} is equivalent to a linear sum assignment problem. This problem is solvable in polynomial time using Hungarian Algorithm \cite{kuhn1955hungarian}, which efficiently finds the permutation $\sigma$ minimizing the total cost $\Tr(\mathbf{\Pi}(\mathbf{\alpha \mathbf{L}_n - \mathbf{q}^\top}\mathbf{c}))$. In our experiments, we instead solve the resulting linear assignment problem using a shortest augmenting path algorithm\cite{Crouse2016assignment}. The polynomial time complexity of the re-ranking algorithm allows for fast refreshment of the user feed, as well as possible implementation of the algorithm on the client side, causing no additional load on the web servers.

\section{Experiments}
\subsection{Validation of Credibility Assignment} \label{sec_ragtests}
To ensure that the re-ranking method improves the credibility of user feeds it is critical for the credibility scores to be accurate. 
An important contributor to the accurate assessment of credibility scores is accurate artificial misinformation labeling we discuss in section \ref{sec:artificiallabels}.
The accuracy of credibility score estimation in the proposed method depends significantly on the performance of the artificial misinformation labeling process, as discussed in Section \ref{sec:artificiallabels}
In this section, we test and compare the method we use with several variants and alternatives to assess their performance.

We test and compare the misinformation labeling performance of seven models on the PolitiFact dataset. We test \textit{GPT-3.5-turbo},\textit{ GPT-4o-mini}, and \textit{GPT-4o} \cite{OpenAI} to represent the baseline performance of OpenAI's LLMs without additional retrieval augmentation. \textit{GPT-4o-mini+AskNews} and \textit{GPT-4o+AskNews} use AskNews' API to query $10$ news articles relevant to the claim, and then use the corresponding LLM to label the claim based on the contextual information from the news articles. A caveat is that the API access in AskNews is only as far back as January $2024$. Thus it is impossible to access news articles that are contemporary with the claims made before $2024$. For these claims, the GPT-4o-mini+AskNews and GPT-4o+AskNews methods default to retrieving the most recent $10$ articles with the highest similarity to the claim. Finally, the \textit{GPT-4o-mini-mix} and \textit{GPT-4o-mix} models use the AskNews' API for only the claims made after $2024$, and default to the base models without retrieval for claims made before $2024$.

PolitiFact scores each claim on a misinformativeness scale ranging from $0$ to $5$, with lower scores corresponding to more misinformative claims. We convert the Politifact scores to binary ground truth labels by thresholding them. That is, we select a threshold value between $0$ and $6$, then label all claims that have a PolitiFact score less than the threshold as misinformative, and label the rest as not misinformative. It is customary in the existing literature to choose the threshold as $3$ \cite{tian2024web}, meaning only the claims with scores $0,1,2$ are considered misinformative. Choosing the threshold as $3$ results in a balanced label split where $600$ claims out of $1165$ are labeled as misinformative. 


\begin{figure}
    \centering
    \includegraphics[width=\linewidth]{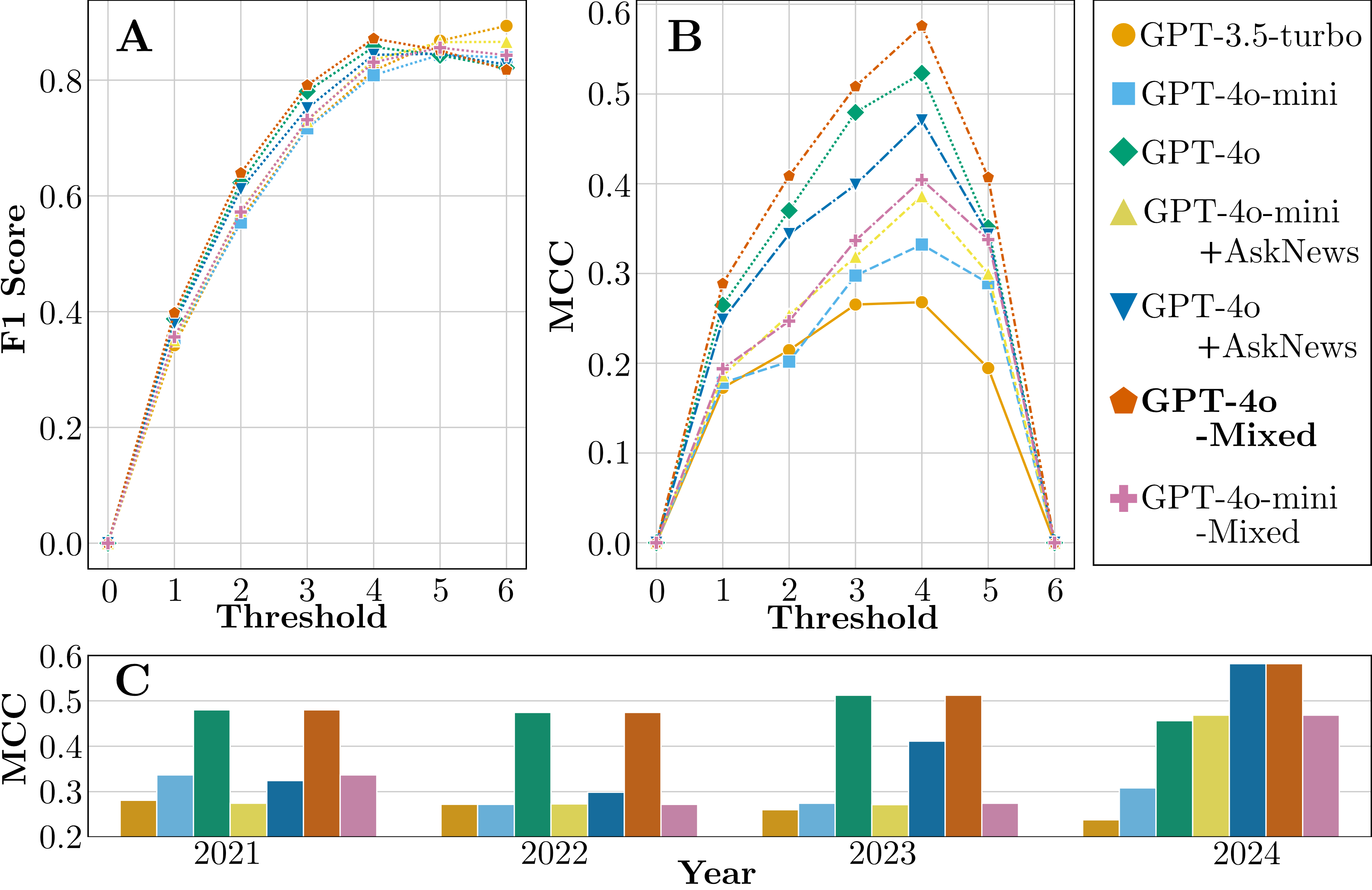}
    \caption{The performance comparison of different artificial credibility assignment methods on Politifact data. A and B show different accuracy metrics across different thresholds for ground truth with (A) F1 Scores and (B) Matthews correlation coefficients (MCC) across different thresholds for ground truth. C shows the Matthews correlation coefficient across years.}
    \label{fig:llm-tests}
\end{figure}

The results in \ref{fig:llm-tests} initially suggest that the base model GPT4o performs better than the retrieval augmented version, GPT-4o+AskNews. This performance difference is mainly due to these tests encompassing the data from the entire range of the dataset, from January $2021$ to October $2024$. This means that the AskNews API does not work on relevant news articles on approximately $76\%$ of the claims in these tests, and feeding recent news by default. 
The absence of contemporary news seems to negatively impact the performance of retrieval-augmented methods. Figure \ref {fig:llm-tests}C visualizes this impact, where the retrieval-augmented methods get a large boost in performance in the year $2024$. When contemporary news sources are available, the retrieval augmentation boosts the performance of GPT4o-mini to that of GPT4o without retrieval, despite the former's smaller size, and GPT-4o+AskNews yields significantly better performance with a $\Delta$MCC of $12\%$ compared to GPT4o when the threshold is set to $3$.

The performance of the models depends heavily on the choice of the threshold value used to assign ground-truth misinformation labels. Figure \ref{fig:llm-tests}A and B show the F1 scores and MCCs of the models for each possible threshold value. The threshold scores directly determine what is considered to be misinformative based on the PolitiFact scores with a threshold $0$ meaning no claim is considered misinformative, and a threshold of $6$ meaning every claim is considered to be misinformative. Though we include the F1 scores for completion, they are uninformative for large threshold values due to the overwhelming abundance of misinformative labels in the ground truth. The MCC provides a more informative comparison between the algorithms. All of the models provide optimal correlation with the ground-truth labels when the threshold is $4$, meaning all claims scored below $4$ by PolitiFact are considered misinformation.



\subsection{Re-ranking Performance}

\begin{figure*}
    \centering
    \includegraphics[width=\linewidth]{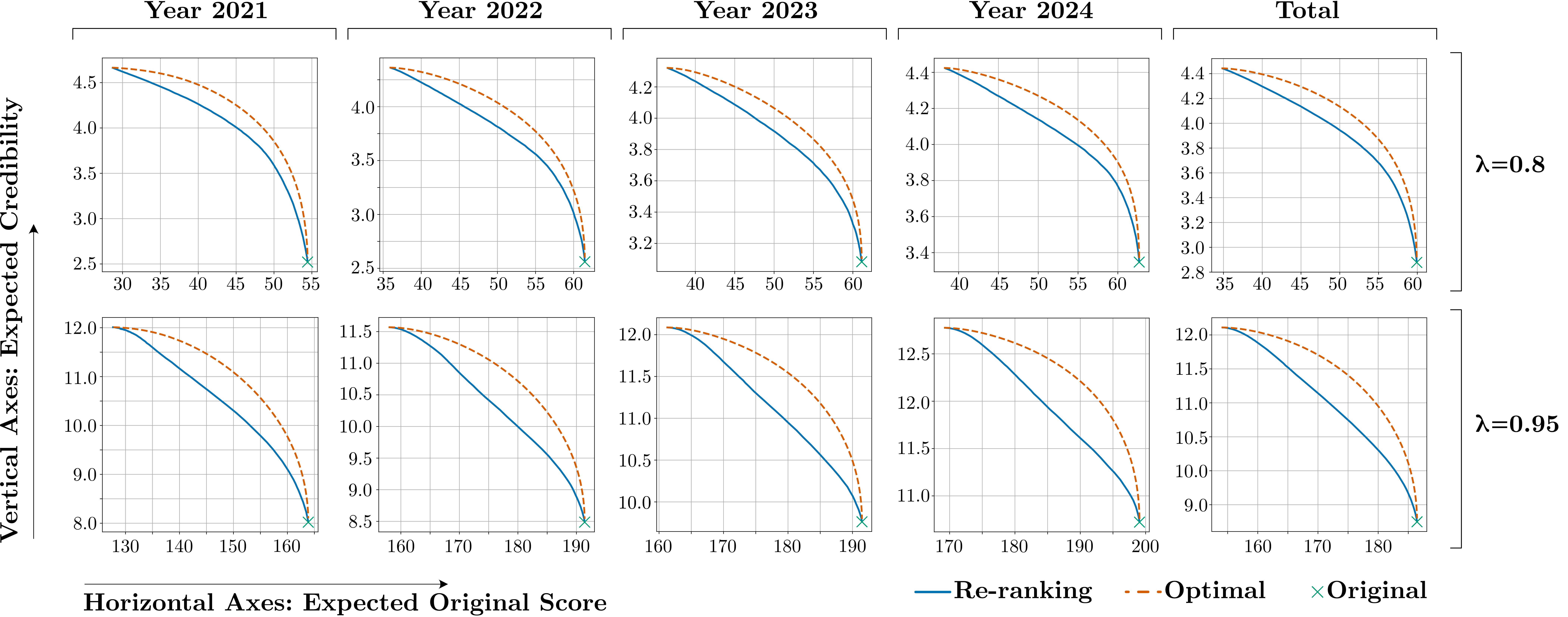}
    \caption{Comparison of Pareto optimal fronts across different years and $\lambda$. Blue solid lines represent the re-ranking method, while red dashed lines represent the theoretical optimal solution to equation \ref{mainprob1} with known original ranking scores. The green crosses in the bottom right corners represent the original ranking. The re-ranking approach obtains a trade-off between original scores and credibility.}
    \label{fig:pareto}
\end{figure*}
To measure the impact of re-ranking, we first generate sample content feeds in canonical ordering. Ideally, we would implement the ranking using the recommendation algorithm X uses to rank user feeds. However, despite the majority of the recommendation system of X being open-source\footnote{Accessible from https://github.com/twitter/the-algorithm}, several factors prohibit such an implementation. Firstly, the recommendation algorithm of X relies on private data that is inaccessible to the public, including the information related to the user for whom the feed is prepared. Secondly, despite the majority of the algorithm being open source, the top-level build files are not, preventing the entire system from being replicated. Thirdly, the X dataset only includes a sample set of posts, and not the full historical database, making any attempt to replicate X's rankings impossible regardless of algorithm availability. 

Instead, we measure the relative impact of the re-ranking by generating canonical rankings using a simple ranking rule based on the number of \textit{likes} on each post. Likes are tags users can attach to posts to show their interest and improve the post ranking score in X's recommendation algorithm. We create the sample rankings by first selecting a random time uniformly from the span of post times in the X community notes dataset. This randomly selected time represents the time at which we generate the feed. Then, we collect the $100$ posts that have the largest post time smaller than the random time we select. This selection imitates choosing the recent posts at the time of the feed generation. Next, we rank the $100$ selected posts based on the logarithm of the number of likes they receive. That is, we define the original score $v(\textrm{post})$ of each post as
\begin{equation}
    v(\textrm{post}) = \log(\textrm{\# of likes of post}),
\end{equation}
and then sort the posts in a descending order of $v$. We sample the top $50$ posts among these $100$ to create a sample content feed of $50$ contents sorted in the descending order of $v$. 

This score assignment provides a canonical ranking and original scores that are both interpretable and independent from the credibility objective. Thus the relative improvement in credibility over this ranking provides a reliable measure for evaluating the re-ranking method. Additionally, having access to the original scores $v$ makes it possible to solve the dual objective optimization problem in equation \ref{mainprob1} directly by sorting the weighted sum of the objectives. This solution provides the theoretical optimal trade-off between maximizing credibility and maximizing the original scores, providing a benchmark to compare with the proposed re-ranking method. The difference between these solutions is that the proposed re-ranking approach does not require the original scores, operating on the canonical rankings directly by solving the surrogate problem in equation \ref{mainprob2}.

We calculate the artificial credibility scores in each post using the model GPT-4o+mix, which has the overall best performance among the models we tested with an F1 score of $79\%$ on the PolitiFact data with threshold $3$. We mix these artificial scores with the average human-generated scores using equation \ref{credscores}, where we choose $k^+ = 10$ and $k^- = 0$ to assign a final credibility label to each post. We then generate a range of Pareto optimal solutions to both equation \ref{mainprob1} through sorting and equation \ref{mainprob2} through re-ranking by sweeping the trade-off parameter $\alpha$ from $2^{-14}$ to $2^6$. We evaluate the solutions according to the expected total original scores and expected total credibility seen by the user interacting with the content feed. We average the results across $500$ generated content feeds for each year to account for the randomness in the data.

Figure \ref{fig:pareto} shows the resulting Pareto optimal fronts for different years and two different user models. $\lambda = 0.8$ represents a user with a short engagement span, which on average sees only $5$ posts in the feed, and $\lambda = 0.95$ represents a user with a long engagement span, which sees on average $20$ posts each time they open their feed. Both the theoretical optimal solution and re-ranking method generate trade-offs ranging from not altering the canonical ranking, therefore only optimizing for the original score, to completely reordering the content feed to optimize solely for credibility. The gap between the optimal and re-ranking indicates the added optimality cost incurred by the re-ranking method due to its requirement to be oblivious to the original scores. 

The yearly variance of the Pareto optimal solutions is low. And despite the slightly different numerical values in expected original score and expected credibility, the ratio-wise gap between the theoretically optimal solutions and re-ranking solutions does not change across years. This is not altogether surprising, as the overall credibility and like distributions of posts remain relatively constant each year. However, it is nevertheless significant that the optimality of the re-ranking method does not depend on the yearly evolution of the topics discussed on the platform and the performance of the algorithm is consistent across time.


In contrast, the performance of the re-ranking algorithm depends heavily on the value for $\lambda$, which determines the average number of posts the user engages with. The re-ranking method achieves near-optimal performance for users who engage with only a few posts. For instance, for $\lambda = 0.8$, the peak difference between the theoretical optimal and re-ranking solutions is less than $4\%$ in both expected credibility and original score objectives. However, as the users start engaging with more posts, the effectiveness of re-ranking diminishes, as apparent from the larger optimality gap between the optimal and re-ranking solutions. The re-ranking algorithm measures the distance to the original rankings using Spearman's footrule distance and penalizes all deviations from the original ranking evenly. In contrast, the optimal solution to equation \ref{mainprob1} tends to penalize deviations from the original ranking proportionally with probabilities $q_p$, which indicate the probability of the user seeing post $p$. This discrepancy becomes more apparent for users who engage with the feed longer, resulting in a larger optimality gap.

\section{Conclusion}

This paper shows the effectiveness of combining re-ranking and retrieval-augmented misinformation detection to increase the credibility of the content seen by social media users. The mixture of human ratings and retrieval-augmented misinformation detection provided a method to measure credibility in a way that is both democratic and informed. The weight of the human-generated and artificial labels is also tunable, increasing the general applicability of the proposed method. The results indicate that re-ranking based on credibility scores can be an effective tool for combating misinformation. Additionally, this approach encourages media outlets to be more truthful to avoid losing their influence. All of the data we use in preparing this paper are publicly available. We do not collect or release any information that is not available to the general public or report any personally identifiable information within the paper.
\bibliography{references}

\end{document}